\documentclass[a4paper,12pt]{article} %
\usepackage{amsmath}
\usepackage{amssymb,amsfonts}
\usepackage[comma,super]{natbib}%
\usepackage{latexsym,rotating}

\begin{document}
%
\title{\textsf{Anisotropic static solutions in modelling highly compact
bodies}}
\author{M. Chaisi\thanks{Permanent address: Department of Mathematics \& Computer Science, National
University of Lesotho, Roma 180, Lesotho.; eMail:
\texttt{m.chaisi@nul.ls}}  \;and S. D. Maharaj\thanks{eMail: \texttt{maharaj@ukzn.ac.za}}\\
School of Mathematical Sciences \\ University of KwaZulu-Natal\\
Durban 4041 \\ South Africa.}

\date{}
\maketitle

\begin{abstract}
Einstein field equations for anisotropic spheres are solved and
exact interior solutions obtained. This paper extends earlier
treatments to include anisotropic models which accommodate a wider
variety of physically viable energy densities. Two classes of
solutions are possible. The first class contains the limiting case
$\mu\propto r^{-2}$ for the energy density which arises in many
astrophysical applications. In the second class the singularity at
the center of the star is not present in the energy density. The
models presented in this paper allow for increasing and decreasing
profiles in the behavior of the energy density.
\end{abstract}
\newpage
\section{Introduction}

Anisotropic spheres in general relativity, in spherically symmetric
spacetimes which are not static, have continued to attract the
attention of researchers since the pioneering work of Bowers and
Liang\cite{BowersLiang}. Some of the interesting recent papers
published in this regard include the works of Dev and Gleiser
\cite{DevGleiser,DevGleiser2}, Herrera \emph{et
al}\cite{HerreraMartin}, Ivanov\cite{Ivanov}, Mak and
Harko\cite{MakHarko2002}, and Sharma and
Mukherjee\cite{SharmaMukherjee2002}. The original reason for
studying relativistic anisotropic spheres was to generate models
that permit redshifts higher than the critical redshift $z_c$ of
isotropic matter.  The analysis of Bondi\cite{bondi} demonstrated
the feasibility and stability of higher redshifts for anisotropic
Newtonian and relativistic stars. Anisotropy cannot be neglected in
stellar clusters and galaxies, in addition to individual stars, as
pointed out by Binney and Tremaine\cite{binneytremaine}, Cuddeford
\cite{cuddeford} and Michie\cite{michie}.  Anisotropy may be an
intrinsic feature of boson stars as suggested by Dev and
Gleiser\cite{DevGleiser}  and Mak and Harko\cite{MakHarko2002}.
Sharma and Mukherjee\cite{SharmaMukherjee2002} considered the
theoretical possibility of anisotropy in strange stars, with
densities greater than neutron stars but less than black holes,
which are bound stable stellar bodies.

In a recent treatment Chaisi and Maharaj\cite{CM1} found an exact
anisotropic solution to the field equations utilizing an algorithm
of Maharaj and Maartens\cite{MaharajMaartens}. This solution has a
simple form: it can be expressed completely in terms of elementary
functions and it has a clear physical interpretation. In this
analysis we extend the Chaisi and Maharaj\cite{CM1} solution to
include anisotropic models which allow for a wide variety of
densities which are physically reasonable. We present four classes
of solution and discuss their physical features. In \S2 we present
the relevant classes of solutions  in the form of tables to ensure
clarity and to highlight the interconnections between the classes.
In \S3 the physical features of our solutions are briefly discussed,
and \S4 contains the discussion.

\section{New classes of solutions}

The Einstein field equations to be integrated are
\begin{eqnarray}
e^{-\lambda} & = & 1-\frac{2m}{r} \label{EFEs2:1} \\
r(r-2m)\nu^\prime & = & p_r r^3+2m \label{EFEs2:2} \\
\left(\mu+p_r\right)\nu^\prime+2p^\prime_r & = &
-\frac{4}{r}\left(p_r-p_\perp\right) \label{EFEs2:3}
\end{eqnarray}
In the above we have defined $ m(r)  =
\frac{1}{2}\int^r_0x^2\mu(x)\mbox{d}x \label{massFun}$ which is the
mass function. The radial pressure $p_r=p+2S/\sqrt{3}$ and the
tangential pressure $p_\perp  = p-S/\sqrt{3}$ are not equal for
anisotropic matter. The magnitude $S$ provides a measure of
anisotropy.

The field equations (\ref{EFEs2:1})-(\ref{EFEs2:3}) were integrated
by Chaisi and Maharaj \cite{CM1} for the energy density
\begin{equation}
\mu
 = \frac{j}{r^2}+k+\ell r^2
 \end{equation}
 where $j,k, \ell$ are constants.
  It is possible to integrate the field
equations
 for other choices of the energy
density which are physically acceptable. In this section we present
three new classes of exact solutions to supplement the class found
earlier.

\vspace{.15in}

\begin{table}[h]\centering
\begin{tabular}{|c|c|}\hline\hline
Case & $\mu\left(r\right)$ \\ \hline\hline I & $\frac{j}{r^2}+k+\ell
r^2$
\\ \hline II & $\frac{1}{r^2}\left(1-a\right)-(p+1)br^{p-2}$
\\ \hline III &
$\frac{1}{r^2}\left(\frac{a-c+(b-d)r^2}{a+br^2}\right)
 -2\frac{d\left(a+br^2\right)-b\left(c+dr^2\right)}{\left(a+br^2\right)^2}$ \\
\hline IV &
$\frac{1}{r^2}\left(\frac{a-1+br^2+cr^4}{a+br^2+cr^4}\right)
+2\frac{b+2cr^2}{\left(a+br^2+cr^4\right)^2}$
\\ \hline \hline
\end{tabular}\caption{\label{mu_table} Energy density functions}
\end{table}

Table~\ref{mu_table} comprises a list of forms for energy densities
that have been studied. The forms of energy density $\mu$ in
Table~\ref{mu_table} were chosen so that the Einstein field
equations could be fully integrated, and the gravitational
potentials and matter variables written in closed form. The
functions chosen for $\mu$ in Table \ref{mu_table} have profiles
which correspond to physically acceptable anisotropic spheres, and
their simple forms facilitate the analysis of the gravitational
potentials and the matter variables. To complete the integration of
the Einstein field equations we also need to make a choice for the
radial pressure $p_r$. Clearly a variety of choices for $p_r$ is
possible; our choice is made on physical grounds. The following is a
list of forms for the radial pressure:

\begin{table}[h]\centering
\begin{tabular}{|c|c|}\hline\hline
Case & $p_r\left(r\right)$ \\ \hline\hline I &
$\frac{C}{1-j}\left(1-j-\frac{k}{3}r^2-\frac{\ell}{5}r^4\right)
\left(1-\frac{r^2}{R^2}\right)^n$
\\ \hline II & $\frac{C}{a}\left(a+br^p\right)\left(1-\frac{r^2}{R^2}\right)^n$
\\ \hline  III &
$\frac{a}{c}C\left(\frac{c+dr^2}{a+br^2}\right)\left(1-\frac{r^2}{R^2}\right)^n$ \\
\hline IV & $\frac{aC}{a+br^2+cr^4}\left(1-\frac{r^2}{R^2}\right)^n$
\\ \hline \hline
\end{tabular}\caption{\label{pr_table} Radial pressure functions}
\end{table}

The forms of $p_r$ selected in Table~\ref{pr_table}  all reduce to
the expression $ p_r = C\left(1-r^2/R^2\right)^n $ with appropriate
choices for the parameters so that the radial pressure is a
monotonically decreasing function. Table \ref{potentials} contains
the gravitational potentials $e^\nu$ and $e^{\lambda}$.
Table~\ref{pperpTable2}  lists the corresponding matter variables:
the energy density $\mu$, the mass function $m$, the radial pressure
$p_r$ and the tangential pressure $p_\perp$, respectively.

We believe that the families of solutions for Cases I-IV presented
in Tables~\ref{potentials}-\ref{pperpTable2} are new solutions to
the field equations, apart from particular special cases, for
relativistic anisotropic matter. Case I contains the Maharaj and
Maartens \cite{MaharajMaartens} solution ($j=\ell=0$) and the
Gokhroo and Mehra \cite{gokhroo} solution ($j=0$); $I_2$ is defined
in terms of elementary functions \cite{CM1}. The exact solutions in
Cases I-IV are amenable to a physical analysis because they have a
simple form, and in all cases the gravitational potentials and
matter variables are given in terms of elementary functions. The
parameters $j$, $k$, and $\ell$ in Case I are constants. The
quantities $a$, $b$, $c$, $d$ and $p$ are also constants in Cases
II-IV. Note that the constant $C$ corresponds to central radial
pressure ($C=p_r(0),n\geq 1$)

\vspace{.25in}

\begin{table}[h] \centering
\begin{tabular}{|c|c|c|} \hline \hline
Case  & $e^\lambda$ & $e^\nu$  \\ \hline \hline I &
$\left(1-j-\frac{k}{3}r^2-\frac{\ell}{5}r^4\right)^{-1}$ &
$Br^{\frac{j}{1-j}}
\exp\left\{\frac{I_2}{1-j}-\frac{CR^2}{2(1-j)(n+1)}
\left(1-\frac{r^2}{R^2}\right)^{n+1}\right\}$ \\
\hline II & $\left(a+br^p\right)^{-1}$ &
$\frac{Br^{\frac{1-a}{a}}}{\left(a+br^p\right)^{\frac{1}{ap}}}
\mbox{exp}\left\{-\frac{CR^2}{2a(n+1)}\left(1-\frac{r^2}{R^2}\right)^{n+1}\right\}$
\\ & $(p>0)$ &\\ \hline
III & $\frac{a+br^2}{c+dr^2}$ &
$Br^{\frac{a-c}{c}}(c+dr^2)^{\frac{bc-ad}{2cd}}
\mbox{exp}\left\{-\frac{aCR^2}{2c(n+1)}\left(1-\frac{r^2}{R^2}\right)^{n+1}\right\}$
\\ \hline IV & $a+br^2+cr^4$ &
$Br^{a-1}\mbox{exp}\left\{-\frac{aCR^2}{2(n+1)}
\left(1-\frac{r^2}{R^2}\right)^{n+1}+\frac{br^2}{2}+\frac{cr^4}{4}\right\}$
\\
\hline\hline
\end{tabular}
\caption{\label{potentials} Gravitational potentials}
\end{table}

\section{\label{sec:phys} Physical features}

The gravitational potentials $e^{\lambda}$ are finite for all Cases
I-IV at the centre $r=0$ and at the boundary $r=R$. The functions
$e^\lambda$ are continuous and well behaved in the interior of the
relativistic star. The gravitational potentials $e^\nu$ for all
Cases I-IV are continuous and well behaved in the interior and
finite at the boundary of the star $r=R$. However we observe from
Table~\ref{potentials} that there is a singularity at the centre
$r=0$ in general for all Cases I-IV in the potential $e^\nu$. The
singularity in $e^\nu$ is not present for specific choices of
parameter values and may be removed by setting
\begin{eqnarray}
  j&=&0, \qquad\mbox{ in Case I,}\label{eq1}
  \\ a&=&1,\qquad \mbox{ in Cases II and IV,} \label{eq2} \\
 a&=&c,\qquad \mbox{ in Case III.}\label{eq3}
\end{eqnarray}
The gravitational potentials in Table~\ref{potentials} have the
advantage of having a simpler analytic form, and they are written in
terms of polynomials, rational and exponential functions.
Consequently the radial and tangential pressures have a simple
analytic representation.

The radial pressure $p_r$ is continuous and well behaved in the
interior. Also $p_r>0$ in the interval $(0,R)$, regular at the
centre $\left(p_r(r=0)=C\right)$, and vanishes at the boundary
$\left(p_r(r=R)=0\right)$ in all four cases. The tangential pressure
$p_\perp$ in the four cases has a singularity at the centre, but is
otherwise well behaved throughout the interior of the star and
 finite at the boundary. Note that the singularity in $p_\perp$ may
be avoided by suitable particular choices for parameter values. In
general the tangential pressure is not zero at the boundary of the
star $\left(p_\perp(r=R)\ne 0\right)$ and does not vanish, as does
the radial pressure $\left(p_r(r=R)= 0\right)$. It is also important
to observe that the magnitude of the stress tensor $ S $ is a
nonzero function in general for all Cases I-IV. Hence this class of
solutions is generally anisotropic and it is not possible for $S$ to
vanish and obtain an isotropic limit.

The energy density $\mu$ for all cases contains the limiting case
\begin{equation}
\mu \propto r^{-2} \label{mupropto}
\end{equation}
as can be  verified directly from Table~\ref{pperpTable2}. It has
been demonstrated that the energy density (\ref{mupropto}) arises in
isothermal spheres for isotropic matter for both Newtonian and
relativistic stars by Saslaw {\em et al} \cite{SaslawMaharaj}. There
are many other examples where the form (\ref{mupropto}) arises in
relativistic astrophysics. Misner and Zapolsky \cite{MisnerZapolsky}
proposed that isotropic solutions (with $k=0=\ell$) model the
physical  configuration of a relativistic Fermi gas for some
particular value of the parameter $j$. Another example, given by Dev
 and Gleiser \cite{DevGleiser} who indicate that for particular values of $j\ne
0$ and $k\ne 0$ ($\ell=0$)  the energy density $\mu$ is appropriate
for modelling the relativistic Fermi gas core immersed in a constant
density background. Consequently our exact solutions in Cases I-IV,
all containing the limiting energy density (\ref{mupropto}), may be
used to model a variety of similar astrophysical situations when the
anisotropy is not negligible.

\begin{figure}[thb]
\vspace{1.8in} \includegraphics{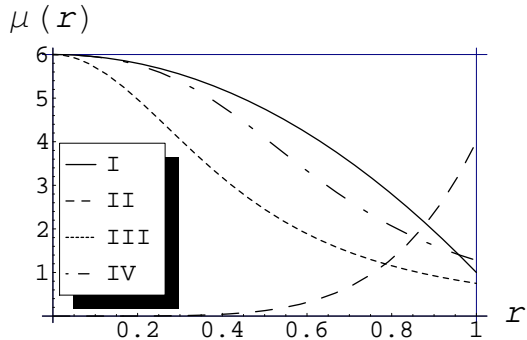} \caption{\label{fig:mu}Energy density
$\mu$ plots with singularities removed}
\end{figure}

It is also important to observe that our approach allows us to
consider energy densities for which (\ref{mupropto}) does not hold.
For this scenario we must utilize the values for the constants in
(\ref{eq1})-(\ref{eq3}). With the help of (\ref{eq1})-(\ref{eq3}) we
observe from Table~\ref{pperpTable2} that the singularity at $r=0$
is removed and $\mu$ is a continuous function throughout the
interval $[0,R]$. Figure~\ref{fig:mu} provides an illustration of
the behaviour of the energy density $\mu$, when
(\ref{eq1})-(\ref{eq3}) is true, for particular chosen values of the
constants. The radial distance is over the interval $0\leq r\leq 1$.
From Figure~\ref{fig:mu} we observe that $\mu$ is a well behaved
function in the interior of the star and has  finite values at the
center and the boundary. Cases I-IV admit both possibilities of
$\mu^\prime<0$ and $\mu^\prime>0$; that is anisotropic stars with
decreasing or increasing energy densities from the centre to the
boundary, respectively, can be studied.

\begin{sidewaystable}
\centering
\begin{tabular}{|c|c|c|c|c|} \hline \hline
Case & $\mu$& $m$ & $p_r$ ($C=p_r(0),n\geq 1$) & $p_\perp$  \\
\hline \hline I & $\frac{j}{r^2}+k+\ell r^2$ &
$\frac{r}{2}\left(j+\frac{k}{3}r^2+\frac{\ell}{5}r^4\right)$ &
$\frac{C}{1-j}\left(1-j-\frac{k}{3}r^2-\frac{\ell}{5}r^4\right)$ &
$p_r+\frac{C}{2\left(1-j\right)}\left(j-\frac{\ell}{5}r^4\right)
\left(1-\frac{r^2}{R^2}\right)^{n}
+\frac{r^2}{2}\left(1-j-\frac{k}{3}r^2-\frac{\ell}{5}r^4\right)^{-1}$\\
&  &  & $\times\left(1-\frac{r^2}{R^2}\right)^n$ &
$\times\left\{\frac{C^2}{2(1-j)^2}\left(1-j-\frac{k}{3}r^2-
\frac{\ell}{5}r^4\right)^2\left(1-\frac{r^2}{R^2}\right)^{2n}\right.$\\
& & & &
$-\frac{2nC}{(1-j)R^2}\left(1-j-\frac{k}{3}r^2-\frac{\ell}{5}r^4
\right)^2\left(1-\frac{r^2}{R^2}\right)^{n-1}$\\
& & & & $\left. +\frac{1}{2r^2}\left(\frac{j}{r^2}+k+\ell
r^2\right)\left(j+\frac{k}{3}r^2+\frac{\ell}{5}r^4\right)\right\}$\\
 \hline
 II & $\frac{1}{r^2}\left(1-a\right)$ & $\frac{r}{2}\left(1-a-br^p
 \right)$  & $\frac{C}{a}\left(a+br^p\right)$ & $p_r+\frac{C}{a}
 \left(\frac{1-a}{2}+b\left(\frac{p}{4}-\frac{1}{2}\right)r^p\right)
 \left(1-\frac{r^2}{R^2}\right)^n +\frac{r^2}{2}\left(a+br^p\right)^{-1}$ \\
 & $-(p+1)br^{p-2}$ & $(p>0)$  & $\times\left(1-\frac{r^2}{R^2}
 \right)^n$ & $\times\left\{\frac{C^2}{2a^2}\left(a+br^p\right)^2
 \left(1-\frac{r^2}{R^2}\right)^{2n}
 -\frac{2nC}{aR^2}\left(a+br^p\right)^{2}\left(1-\frac{r^2}{R^2}
 \right)^{n-1}\right.$
 \\ & & & & $\left.
 +\frac{1}{2r^4}\left(1-a-br^p\right)\left(1-a-b(p+1)r^p\right)
 \right\}$ \\ \hline
 III & $\frac{1}{r^2}\left(\frac{a-c+(b-d)r^2}{a+br^2}\right)$
 & $\frac{r}{2}\left(1-\frac{c+dr^2}{a+br^2}\right)$
 & $\frac{a}{c}C\left(\frac{c+dr^2}{a+br^2}\right)$ & $p_r+
 \frac{r}{4}\left\{\frac{a-c+(b-d)r^2}{r^2(a+br^2)}-\frac{2
 \left(ad-bc\right)}{(a+br^2)^2}+\frac{aC}{c}\left(\frac{c+dr^2}{a+br^2}
 \right)\left(1-\frac{r^2}{R^2}\right)^n\right\}$ \\
 & $-\frac{2\left(ad-bc\right)}{\left(a+br^2\right)^2}$ &
  & $\times\left(1-\frac{r^2}{R^2}\right)^n$ & $\times
  \left\{\frac{aCr}{c}\left(1-\frac{r^2}{R^2}\right)^n+
  \frac{a-c+(b-d)r^2}{r(c+dr^2)}\right\}$
 \\ & & & & $+\frac{aCr^2}{c}\frac{ad-bc}{(a+br^2)^2}
 \left(1-\frac{r^2}{R^2}\right)^n -\frac{anCr^2}{cR^2}
 \left(\frac{c+dr^2}{a+br^2}\right)\left(1-\frac{r^2}{R^2}
 \right)^{n-1}$ \\ \hline
  IV & $\frac{1}{r^2}\left(\frac{a-1+br^2+cr^4}{a+br^2+cr^4}\right)$
 & $\frac{r}{2}\left(1-\frac{1}{a+br^2+cr^4}\right)$
 & $\frac{aC}{a+br^2+cr^4}$ &$p_r+\frac{r}{4}\left\{
 \frac{a-1+br^2+cr^4}{r^2(a+br^2+cr^4)}+\frac{2
 \left(b+2cr^2\right)}{(a+br^2+cr^4)^2}+\frac{aC}{a+br^2+cr^4}
 \left(1-\frac{r^2}{R^2}\right)^n\right\}$ \\
 & $+\frac{2\left(b+2cr^2\right)}{\left(a+br^2+cr^4
 \right)^2}$ &  & $\times \left(1-\frac{r^2}{R^2}
 \right)^n$ & $\times
\left\{aCr\left(1-\frac{r^2}{R^2}\right)^n+
\frac{1}{r}(a-1+br^2+cr^4)\right\}$
\\ & & &  & $-aCr^2\frac{b+2cr^2}{(a+br^2+cr^4)^2}
\left(1-\frac{r^2}{R^2}\right)^n -\frac{aCnr^2}{R^2
(a+br^2+cr^4)}\left(1-\frac{r^2}{R^2}\right)^{n-1}$ \\
 \hline \hline
\end{tabular}
\caption{\label{pperpTable2} Matter variables}
\end{sidewaystable}

\pagebreak

\section{Discussion\label{discussion}}

We have presented four new classes of solution which model
anisotropic stars. These solutions may be broadly divided into two
categories in terms of the behaviour of the energy density $\mu$ at
the centre $r=0$: the first category contains the limiting case
$\mu\propto r^{-2}$, and in the  second category  the singularity at
the centre of the star is not present. The solutions found permit
$\mu^\prime<0$ and $\mu^\prime>0$ which allow for decreasing and
increasing energy densities as we move from the centre to the
boundary of the star.  Note that the solutions presented in this
paper have the feature that $p_r\ne p_\perp$ in general so that the
anisotropy factor $S\ne 0$; consequently our solutions do not have
an isotropic limit.

We make three observations relating to the physical reasonableness
of our model. Firstly, the vanishing of the pressure at the boundary
is a consequence of the first and second fundamental forms; there is
no restriction placed on the tangential pressure which may be
nonvanishing. This feature is also evident in the solutions of
Chaisi and Maharaj \cite{CM1}, Maharaj and Maartens
\cite{MaharajMaartens} and Gokhroo and Mehra \cite{gokhroo}.
Secondly, the solutions in this paper do not satisfy an equation of
state. For a realistic astrophysical matter distribution we should
impose a barotropic equation of state relating the pressure and
energy density; this will be pursued in future work. Thirdly, we
observe that for realistic models the space derivatives of the
matter field should be negative. This will have the effect of
restricting the arbitrary constants in the energy density; for
example this would place restrictions on the relative magnitudes of
$j,k, \ell$ of Case I.

\section*{Acknowledgements}

SDM and MC thank the National Research Foundation of South Africa
for financial support. MC is grateful to the University of
KwaZulu-Natal for a scholarship. We are thankful to the referee for
insightful comments.

%

%
%
\end{document}